\documentclass[12pt]{article} 
\usepackage{latexsym,cite,subeqn,graphics}
\input epsfig.sty 
\input amssym.def    
\input amssym.tex  
\makeatletter   
\renewcommand{\theequation}{\thesection.\arabic{equation}} 
\@addtoreset{equation}{section} 
\makeatother  
\font\msym=msbm10 
 
\def\Real{{\mathop{\hbox{\msym \char  '122}}}}

\def\F{{\cal F}} 
\def\tr{\mbox{tr}}  
\def\Tr{\mbox{Tr}}
\def\Det{\mbox{Det}}
\def\L{{\cal L}}

\begin{document}  
\begin{titlepage}
\title{\vskip -60pt
{\small
\begin{flushright} 
DAMTP 98-157\\
APCTP-1999004\\
hep-th/9902081 
\end{flushright}}
\vskip 45pt
A Study of a Non-Abelian Generalization \\
of the  Born-Infeld Action}
\vspace{4.0cm}
\author{\\
\\Jeong-Hyuck Park\thanks{E-mail address:\,J.H.Park@damtp.cam.ac.uk}}
\date{}
\maketitle
\vspace{-1.0cm}
\begin{center}
\textit{Asia Pacific Center for Theoretical Physics}\\
\textit{207-43 Cheongryangri-dong Dongdaemun-gu}\\
\textit{Seoul 130-012, Korea}
\end{center}
\vspace{3.3cm}
\begin{abstract}
A new type of non-Abelian generalization of  the Born-Infeld action is
proposed, in which the spacetime indices and group indices are
combined.  The action is manifestly Lorentz and gauge invariant. 
In its power expansion, the lowest order term is   
the Yang-Mills action and  the second  term corresponds to the 
bosonic stringy correction to this action.  Solutions of 
the Euler-Lagrange equation for the  
$\mbox{SU}(2)$ case are considered and 
we show that there exists an instanton-like  solution   
which has winding number one and  finite action. 
\end{abstract}
\thispagestyle{empty}
\end{titlepage}
\newpage

\section{Introduction}
The Born-Infeld action~\cite{Born-Infeld} has been of renewed
attention  for  reasons connected with 
M/string theory, but perhaps interest has also been enhanced by 
its  compact and elegant form. There have been several
proposals for the non-Abelian generalization of the  Born-Infeld 
action by taking either a trace~\cite{JPA143059}, an 
anti-symmetrized trace~\cite{NPB330151} or a symmetrized
trace~\cite{NPB50141} of the group indices  after formally
expanding the ordinary Born-Infeld Lagrangian with the non-Abelian
gauge fields substituted.  In particular,   the 
anti-symmetrized trace structure generates only odd powers of the field
strength, $F$, while the symmetrized trace structure gives only even
powers.   Bosonic stringy corrections to the Yang-Mills
action give a $F^{3}$ term at the lowest order~\cite{NPB81118} while 
for superstring theory  the cubic term is
absent~\cite{NPB276391,9801127}.  
However, the issue as to what is the natural  non-Abelian generalization of 
the Born-Infeld action does not seem to have been resolved fully 
yet~\cite{9804180}.
\newline
In this paper, we propose a different  non-Abelian generalization of 
the Born-Infeld action for $\mbox{SU}(N)$ gauge fields, where 
the spacetime indices and group indices are combined together. 
The action is manifestly Lorentz and gauge invariant. 
In its power expansion, the lowest order term is   
the Yang-Mills action and  the second  term corresponds to the 
bosonic stringy correction to this action.  Solutions of the  
Euler-Lagrange equation for  the $\mbox{SU}(2)$ case are considered\footnote{While this work was being completed, monopole solutions for the symmetrized non-Abelian Born-Infeld action were discussed in \cite{9901073}.} and
we show that there exists an instanton-like  solution   
which has winding number one and  finite action. 

\section{Non-Abelian Born-Infeld Lagrangian}
The generators of the $\mbox{SU}(N)$ group in the 
fundamental representation,  
$T^{a}=-T^{a}{}^{\dagger},\,\tr\,T^{a}=0,\,
a=1,2,\cdots,N^{2}-1$, may be chosen conventionally to satisfy
\begin{equation}
T^{a}T^{b}=-\textstyle{\frac{1}{2N}}
\delta^{ab}+\textstyle{\frac{1}{2}}(id^{abc}+f^{abc})T^{c}
\label{TT}
\end{equation}
where $d^{abc},\,f^{abc}$ are real and totally symmetric, 
anti-symmetric over the indices, $a,b,c$.  \newline
Eq.(\ref{TT})  implies
\begin{equation}
\tr(T^{a}T^{b})=-\textstyle{\frac{1}{2}}\delta^{ab}
\label{Norm}
\end{equation}
and a completeness relation 
follows from the fact that $\{1,T^{a}\}$ forms a basis of 
$N\times N$ matrices
\begin{equation}
\sum_{a}T^{a}_{\alpha\beta}T^{a}_{\gamma\delta}
=\textstyle{\frac{1}{2N}}\delta_{\alpha\beta}
\delta_{\gamma\delta}
-\textstyle{\frac{1}{2}}\delta_{\beta\gamma}\delta_{\alpha\delta}
\label{complete}
\end{equation}
The gauge field, $A_{\mu}$, is Lie algebra valued
\begin{equation}
A_{\mu}=A^{a}_{\mu}T^{a}
\end{equation}
and the field strength is given by
\begin{equation}
F_{\mu\nu}=\partial_{\mu}A_{\nu}-\partial_{\nu}A_{\mu}+[A_{\mu},A_{\nu}]
\end{equation}
In this paper, we regard the field strength, $F$,
as a $Nd\times Nd$ matrix acting on\newline 
$\Real^{N}_{\rm{gauge}}\otimes\Real^{d}_{\rm{spacetime}}$, 
where $d$ is the spacetime dimension, 
 by pairing the spacetime index, $\mu$, and 
group index, $\alpha$, together
\begin{equation}
F_{\mu\alpha,\,\nu\beta}\equiv F^{a}_{\mu\nu}T^{a}_{\alpha\beta}
\end{equation}
We also expand the spacetime metric, $g_{\mu\nu}$, to a $Nd\times Nd$
matrix
\begin{equation}
g_{\mu\nu}\longrightarrow G_{\mu\alpha,\,\nu\beta}\equiv 
g_{\mu\nu}\delta_{\alpha\beta}
\end{equation}
The non-Abelian Yang-Mills kinetic term is then
\begin{equation}                      
-\textstyle{\frac{1}{4}}F^{a}_{\mu\nu}F^{a\mu\nu}
=-\textstyle{\frac{1}{2}}\Tr\F^{2}
\label{YM}
\end{equation}
where $\F$ is a $Nd\times Nd$ matrix given by
\begin{equation}
\F\equiv G^{-1}F
\end{equation} 
$\Tr$ and $\Det$ indicate 
 the trace and  determinant of $Nd\times Nd$ matrices.\newline
The proposed  non-Abelian generalization of the Born-Infeld    
Lagrangian   is then 
\begin{equation}
\begin{array}{ll}
\L&=2N\kappa^{-2}\left(
\left(\Det(G+\kappa F)\right)^{\frac{1}{2N}}-\sqrt{|g|}\right)\\
{}&{}\\
{}&=2N\kappa^{-2}\sqrt{|g|}\left(\mbox{exp}
\left(\sum^{\infty}_{n=1}\,\frac{(-1)^{n-1}}{n}
\textstyle{\frac{1}{2N}}\kappa^{n}\Tr\F^{n}\right)-1\right)\\
{}&{}\\
{}&=\sqrt{|g|}\left(-\frac{1}{2}\Tr\F^{2}+\frac{1}{3}\kappa\Tr\F^{3}
-\frac{1}{4}\kappa^{2}\Tr\F^{4}+\frac{1}{16N}\kappa^{2}(\Tr\F)^{2}
+\cdots\right)
\end{array}
\label{NBI}
\end{equation}
where $\kappa$ is a coupling constant.  
In the second line we have used the identity,\newline 
$\Det M=\mbox{exp}(\Tr\ln M)$.\newline
The Lagrangian,~$\L$,~(\ref{NBI})  is clearly real 
and  gauge invariant.\footnote{Note that $G,F$ are hermitian. }
The lowest order term or  $\kappa\rightarrow 0$ limit   
corresponds to the ordinary 
Yang-Mills action~(\ref{YM}).\footnote{We also note 
that the cubic term is identical  with 
the bosonic stringy correction to the 
Yang-Mills action~\cite{NPB81118}.}\newline
From eq.(\ref{TT}) we can calculate the traces appearing in
eq.(\ref{NBI})
\begin{equation}
\begin{array}{ll}
\L=&\sqrt{|g|}\left[-\textstyle{\frac{1}{4}}F^{a}_{\mu\nu}F^{a\mu\nu}-
\textstyle{\frac{1}{12}}\kappa f^{abc}F^{a\mu}{}_{\nu}F^{b\nu}{}_{\lambda}
F^{c\lambda}{}_{\mu}\right.\\
{}&{}\\
{}&\,\,+\textstyle{\frac{1}{64N}}\kappa^{2}\left(
(F^{a}_{\mu\nu}F^{a\mu\nu})^{2}-(4\delta^{ab}\delta^{cd}+2Nd^{abe}d^{cde}
-2Nf^{abe}f^{cde})F^{a\mu}{}_{\nu}F^{b\nu}{}_{\lambda}
F^{c\lambda}{}_{\rho}F^{d\rho}{}_{\mu}\right)\\
{}&{}\\
{}&\,\,\left.+\cdots\,\right]
\end{array}
\end{equation}
From now on we restrict to flat spacetime with the metric $\eta$.\newline
If we define\footnote{We define 
$\frac{\partial~~~}{\partial 
F^{a}_{\mu\nu}}J^{b\lambda\rho}F^{b}_{\lambda\rho}=\frac{1}{2}(J^{a\mu\nu}-J^{a\nu\mu})$.}
\begin{equation}
\begin{array}{ll}
\Pi^{a\mu\nu}&=-\Pi^{a\nu\mu}\\
{}&{}\\
{}&=\displaystyle{-2
\frac{\partial \L(F)}{\partial
F^{a}_{\mu\nu}}}=\kappa^{-1}(\Det(G+\kappa
F))^{\frac{1}{2N}}\left((G+\kappa F)^{-1}{}^{\mu\alpha,\,\nu\beta}-
(G+\kappa F)^{-1}{}^{\nu\alpha,\,\mu\beta}\right)T^{a}_{\beta\alpha}\\
{}&{}\\
{}&=F^{a}{}^{\mu\nu}-\textstyle{\frac{1}{2}}\kappa f^{abc}
F^{b}{}^{\mu\lambda}F^{c}{}_{\lambda}{}^{\nu}+\cdots
\end{array}
\label{Pi}
\end{equation}
then the equation of motion is 
\begin{equation}
\begin{array}{cc}
0=D_{\mu}\Pi^{\mu\nu}=\partial_{\mu}\Pi^{\mu\nu}+[A_{\mu},\Pi^{\mu\nu}]
~~~~&~~~~\Pi^{\mu\nu}=\Pi^{a\mu\nu}T^{a}
\end{array}
\label{eqm}
\end{equation}
The energy-momentum tensor may be given either by Noether procedure
\begin{equation}
\begin{array}{ll}
T_{N}^{\mu\nu}&=\displaystyle{
\frac{\partial{\cal L}}{\partial A^{a}_{\lambda,\mu}}
A^{a,\nu}_{\lambda}-\eta^{\mu\nu}{\cal L}}\\
{}&{}\\
{}&=-\Pi^{a\mu\lambda}A^{a,\nu}_{\lambda}-\eta^{\mu\nu}{\cal L}
\end{array}
\end{equation}
or by taking variation with respect to the metric 
\begin{equation}
\begin{array}{ll}
T_{g}^{\mu\nu}&=\displaystyle{-\frac{2}{\sqrt{|g|}}
\frac{\partial{\cal L}}{\partial g_{\mu\nu}}|_{g=\eta}}\\
{}&{}\\
{}&=-\kappa^{-2}
\left(\Det(G+\kappa F)\right)^{\frac{1}{2N}}
\displaystyle{\left((G+\kappa F)^{-1}{}^{\mu\alpha,\,\nu\alpha}+
(G+\kappa F)^{-1}{}^{\nu\alpha,\,\mu\alpha}\right)
{}|_{g=\eta}}+
2N\kappa^{-2}\eta^{\mu\nu}\\
{}&{}\\
{}&=\Pi^{a\mu\lambda}F^{a}_{\lambda}{}^{\nu}-\eta^{\mu\nu}{\cal L}
\end{array}
\label{Tg}
\end{equation}
The last equality in eq.(\ref{Tg}) comes from observing   
\begin{equation}
(G+\kappa F)^{-1}=G^{-1}-\kappa (G+\kappa F)^{-1}FG^{-1}
=G^{-1}-\kappa G^{-1}F(G+\kappa F)^{-1}
\end{equation}
The difference between $T^{\mu\nu}_{N}$ and $T^{\mu\nu}_{g}$ is with 
eq.(\ref{eqm}) 
\begin{equation}
T^{\mu\nu}_{N}-T^{\mu\nu}_{g}=\partial_{\lambda}(\Pi^{a\lambda\mu}A^{a\nu})
\end{equation}
which is consistent with the energy-momentum tensor conservation
\begin{equation}
\partial_{\mu}T_{N}^{\mu\nu}=\partial_{\mu}T_{g}^{\mu\nu}=0
\end{equation}

\section{$\mbox{SU}(2)$ Action in  Four-dimensional  Euclidean Space}
We now consider the  gauge group  $\mbox{SU}(2)$ and also restrict to
four-dimensional 
Euclidean space, $x^{0}\rightarrow -ix^{4}$, 
$\eta_{\mu\nu}=\delta_{\mu\nu}$ so that upper and
lower Lorentz indices may be identified.  We  take 
$T^{a}=-\frac{i}{2}\sigma^{a}$, $f^{abc}=\epsilon^{abc}$, where 
$\sigma^{a},\,a=1,2,3$ are Pauli matrices. \newline
If we define
$E_{\mu\alpha,\,\nu\beta}=\delta_{\mu\nu}\epsilon_{\alpha\beta}$  with
a  $2\times 2$ anti-symmetric matrix, $\epsilon$, then from
$\epsilon\sigma^{a}\epsilon=(\sigma^{a})^{t}$, $E(G+\kappa F)$ 
is anti-symmetric so that
\begin{equation}
\mbox{Det}(G+\kappa F)=(\mbox{Pfaffian}(EG+\kappa EF))^{2}
\end{equation}
Using this 
for the $\mbox{SU}(2)$ case, the Lagrangian~(\ref{NBI}),
$\L\rightarrow -\L_{\rm{E}}$, becomes 
\begin{equation}
\begin{array}{ll}
\L_{\rm {E}}&=4\kappa^{-2}\left[1-\left(1
-\textstyle{\frac{1}{4}}\kappa^{2}\Tr\F^{2}
+\textstyle{\frac{1}{6}}\kappa^{3}\Tr\F^{3}
-\textstyle{\frac{1}{8}}\kappa^{4}\Tr\F^{4}
+\textstyle{\frac{1}{32}}\kappa^{4}(\Tr\F^{2})^{2}
\right)^{\frac{1}{2}}\right]\\
{}&{}\\
{}&=4\kappa^{-2}\left[1-\left(
1-\textstyle{\frac{1}{8}}\kappa^{2}F^{a}_{\mu\nu}F^{a}_{\mu\nu}
-\textstyle{\frac{1}{24}}\kappa^{3}
\epsilon^{abc}F^{a}_{\mu\nu}F^{b}_{\nu\lambda}F^{c}_{\lambda\mu}+
\textstyle{\frac{1}{128}}\kappa^{4}
(F^{a}_{\mu\nu}F^{a}_{\mu\nu})^{2}\right.\right.\\
{}&{}\\
{}&~~~~~~\,\,\,\,\,\,\,\,\,\,\,\,\left.\left.-\textstyle{\frac{1}{32}}
\kappa^{4}
F^{a}_{\mu\nu}F^{a}_{\nu\lambda}F^{b}_{\lambda\rho}F^{b}_{\rho\mu}+
\textstyle{\frac{1}{64}}\kappa^{4}
F^{a}_{\mu\nu}F^{b}_{\nu\lambda}F^{a}_{\lambda\rho}F^{b}_{\rho\mu}
\right)^{\frac{1}{2}}\right]
\end{array}
\label{su2action}
\end{equation} 
Now we seek an $\mbox{SO}(4)$ invariant 
solution in four-dimensional Euclidean space by adopting the ansatz of
\cite{PLB5985}
\begin{equation}
\begin{array}{cc}
A_{\mu}(x)=f(r)g(x)^{-1}\partial_{\mu}g(x)~~~~
&~~~~g(x)=\displaystyle{\frac{1}{r}}
(x_{4}-ix_{i}
\sigma^{i})\in\mbox{SU}(2)
\end{array}
\label{ansatz}
\end{equation}
where $r=\sqrt{{\rm x}^{2}+x_{4}^{2}}$.\newline
After some calculation, we get
\begin{equation}
\begin{array}{l}
F_{\mu\nu}=\left(X
(\delta_{\mu\lambda}\delta_{\nu\rho}-\delta_{\mu\rho}\delta_{\nu\lambda})+
Y\epsilon_{\mu\nu\lambda\rho}\right)x_{\lambda}g^{-1}\partial_{\rho}g\\
{}\\
{{}^{\ast} F}_{\mu\nu}=\textstyle{\frac{1}{2}}\epsilon_{\mu\nu\lambda\rho}
F_{\lambda\rho}=
\left(Y(\delta_{\mu\lambda}\delta_{\nu\rho}-\delta_{\mu\rho}\delta_{\nu\lambda})+X\epsilon_{\mu\nu\lambda\rho}\right)
x_{\lambda}g^{-1}\partial_{\rho}g
\end{array}
\end{equation}
and
\begin{equation}
\Pi_{\mu\nu}=(1-2\kappa Y-3\kappa^{2} X^{2})^{-\frac{1}{2}}
(F_{\mu\nu}+\kappa X{{}^{\ast} F}_{\mu\nu})
\end{equation}
where
\begin{equation}
\begin{array}{cc}
X=\displaystyle{\frac{f^{\prime}}{r}}~~~~&
~~~~Y=\displaystyle{\frac{2}{r^{2}}}(f-1)f
\end{array}
\end{equation}
Using the Bianchi identity
\begin{equation}
D_{\mu}{{}^{\ast} F}_{\mu\nu}=0
\label{Bianchi}
\end{equation}
and 
\begin{equation}
D_{\mu}F_{\mu\nu}=(f^{\prime\prime}+X+2Y-4fY)g^{-1}\partial_{\nu}g
\end{equation}
the equation of motion~(\ref{eqm}) gives a 
second order differential equation
\begin{equation}
0=f^{\prime\prime}(1+\kappa Y)(1-2\kappa Y)
+2(2f-1)(\kappa X^{2}+4\kappa^{2}X^{2}Y-Y+2\kappa Y^{2})
+X(1-5\kappa Y-6\kappa^{2} X^{2})
\label{eqmsu2}
\end{equation}
With the ansatz~(\ref{ansatz}) the action becomes
\begin{subeqnarray}
\label{actionS}
&{\cal S}_{\rm{E}}
=2\pi^2\int_{0}^{\infty}{}{\rm
d}r~r^{3}~\L_{\rm{E}}(r,\kappa)&\label{actiona}\\
{}\nonumber\\
&\L_{\rm{E}}(r,\kappa)=
4\kappa^{-2}\left(1-(1+\kappa Y)
(1-2\kappa Y-3\kappa^{2}X^{2})^{\frac{1}{2}}\right)&
\end{subeqnarray}
One may verify that the equation of motion for this
one-dimensional action is identical to eq.(\ref{eqmsu2}). \newline
If we denote the $k$ dependence of the solution explicitly as 
$f(r,\kappa)$ then from eq.(\ref{eqmsu2}) we get the following   scale 
property 
\begin{equation}
f(r,\kappa)=f(r/\sqrt{\kappa},1)
\label{scale}
\end{equation}
In fact by letting $r/\sqrt{\kappa}\rightarrow r$ it is easy to see
that $\kappa$ can be removed from the action so that effectively we
may take $\kappa=1$.\newline
We have not been able to obtain an analytic solution of
eq.(\ref{eqmsu2}) for $\kappa\neq 0$. However,  it is 
easy to see that there are  three stationary solutions\footnote{
See the appendix for  some analysis of the linearized version of  
eq.(\ref{eqmsu2}) around the stationary solutions.}   
\begin{equation}
\begin{array}{ccc}
f(r)=\textstyle{\frac{1}{2}},~~~~&~~~~
f(r)=0,~~~~&~~~~f(r)=1
\end{array}
\end{equation}
The  action~(\ref{actiona}) is  infinity for $f=\frac{1}{2}$ and
zero for $f=0,1$. \newline
We define an instanton-like 
solution   as one which interpolates between $f(0)=0$ and
$f(r)\rightarrow 1$ as $r\rightarrow\infty$. 
Eq.(\ref{eqmsu2}) is invariant under  
$f(r)\rightarrow {1-f(r)}$.  This relates the 
instanton-like solution  to the   anti-instanton-like solution, 
$f_{\rm{a}}(r)=1-f_{\rm{i}}(r)$.\newline
The same  situation occurs for 
the Yang-Mills action, where the equation of motion is given by 
eq.(\ref{eqmsu2}) with $\kappa=0$
\begin{equation}
0=f^{\prime\prime}{}r^{2}+f^{\prime}{}r-4(2f-1)(f-1)f
\end{equation}
The instanton solution for the Yang-Mills action is
of the exact form
\begin{equation}   
f_{YM}(r)=\displaystyle{\frac{r^{2}}{r^{2}+a}}
\label{exactinstanton} 
\end{equation}
which may be obtained by solving $X=-Y$.\newline
Now if we write $f$ for large $r$ as 
\begin{equation}
\begin{array}{cc}
f(r)=\lambda+\sum^{\infty}_{n=0}\,a_{n}r^{\alpha-n}~~~~~&~~\alpha <0
\end{array}
\end{equation}
and substitute this expression into eq.(\ref{eqmsu2}) then we get
$\alpha=-2,\,\lambda=0,1$. Choosing $\lambda=1$ and $\kappa=1$ we get 
an instanton-like solution 
\begin{equation}
f_{\rm{i}}(r)=\displaystyle{
1-a\frac{1}{r^{2}}+a^{2}\frac{1}{r^{4}}
-(a^{3}+a^{2})\frac{1}{r^{6}}+\cdots}
\label{instanton}
\end{equation}
where $a$ is a free parameter.\newline
A similar analysis can be done for small $r$ and gives
\begin{equation}
f^{\rm{i}}(r)
=br^{2}-\textstyle{\frac{3+20b}{3+12b}}b^{2}r^{4}+
\textstyle{\frac{9+153b+618b^{2}-464b^{3}
-3584b^{4}}{9(1-2b)(1+4b)^{3}}}
b^{3}r^{6}+\cdots
\label{instanton2}
\end{equation}
where $b$ is a free parameter. \newline
If  eqs.(\ref{instanton},\,\ref{instanton2}) are well defined 
for  $r\geq s,\,r\leq s$ respectively and   there exist constants    
$a,b$ such that $f^{\rm{i}}(s)=f_{\rm{i}}(s),\,
f^{\rm{i}\prime}(s)=f^{\prime}_{\rm{i}}(s)$,  then the 
instanton/anti-instanton-like solutions are well defined  
globally.\footnote{In consequence,
there will be  no free parameter left contrary to 
the ordinary instanton solutions, where conformal symmetry of the
Yang-Mills action allows a free parameter. 
However, we have not been able to get  the exact values of $a,b$.}  \newline
With   initial conditions, $f^{\prime}(0)=0,\,f(0)=0$,   
some numerical analysis\footnote{We 
put $\kappa=1$ for the numerical analysis, Fig.~1,\,2,\,3.} 
support this instanton-like  
solution:\newline
\newline
\begin{tabular}{c}
$f(r)$~~~~~~~~~~~~~~~~~~~~~~~~~~~~~~~~~~~~~~~~~~~~~~~~~~~~~~~~~~~~~~~~~~~~~~~\\
~~~~~~~\resizebox{12.0cm}{5.0cm}{\includegraphics{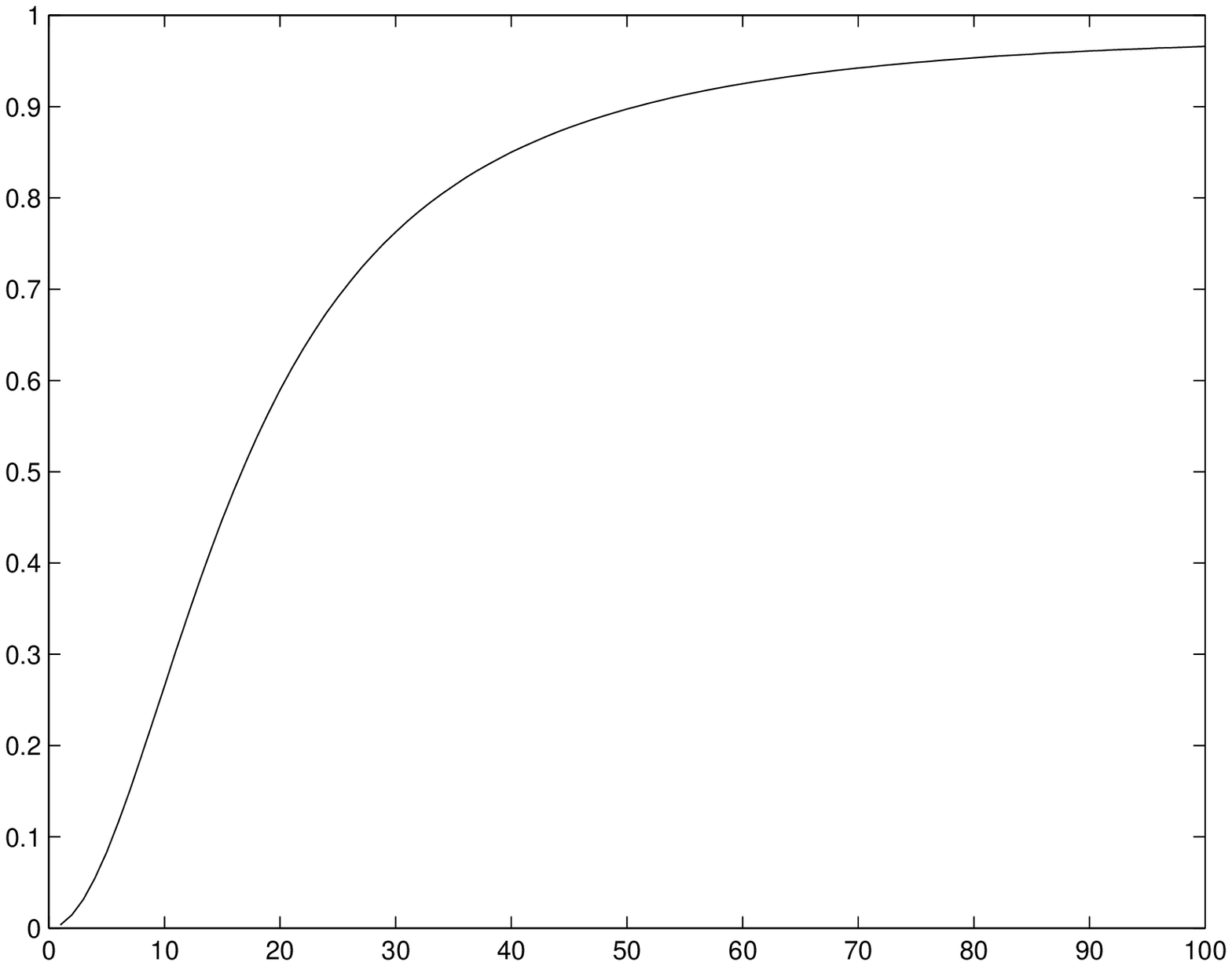}}r\\
~~~~~~~Fig.~1. \textsc{Instanton-Like}
\end{tabular}
\newline
\newline
\newline
As $r\rightarrow\infty$, we have $f\rightarrow 1$ for the instanton-like
solution~(\ref{instanton}) and hence 
\begin{equation}
\begin{array}{cc}
A_{\mu}\rightarrow
g^{-1}\partial_{\mu}g,~~~~&~~~~F_{\mu\nu}\rightarrow 0
\end{array}
\end{equation}
In general this condition is sufficient for 
the integral of the Chern-Pontryagin density, 
${{}^{\ast} F}^{a}_{\mu\nu}F^{a}_{\mu\nu}$, to 
be identical with the topological  Cartan-Maurer form, since
\begin{equation}
{{}^{\ast} F}^{a}_{\mu\nu}F^{a}_{\mu\nu}=\partial_{\mu}\mbox{tr}(
\textstyle{\frac{4}{3}}\epsilon_{\mu\nu\lambda\rho}{}
A_{\nu}A_{\lambda}A_{\rho}-4A_{\nu}{}{{}^{\ast}F}_{\mu\nu})
\end{equation}
and the first term gives the Cartan-Maurer form 
as a surface integral   
which depends on only the value of $f$ at infinity while the second  
term vanishes\,\cite{weinberg23}. 
Thus our instanton-like solution has the same winding number as the 
ordinary instanton solution in the Yang-Mills action, which is one 
\begin{equation}
-\textstyle{\frac{1}{32\pi^{2}}}
\int {\rm d}^{4}x~{{}^{\ast} F}^{a}_{\mu\nu}F^{a}_{\mu\nu}=1
\end{equation}
A similar analysis can be done for the anti-instanton-like  
solution, where $A_{\mu}(x)$ has the same singularity
at the origin  as  
the ordinary anti-instanton solution of the Yang-Mills theory. Hence the
winding number is minus one.\newline
Some numerical analysis also support this result.  Furthermore the
action, ${\cal S}$, seems\footnote{We do not have any rigorous
proof of it.} to have the exact value, $8\pi^2$.  
\newline
\begin{tabular}{ll}
$r^{3}{\cal L}_{\rm{E}}(r)$
&~~~~~${\cal S}_{\rm{E}}/8\pi^{2}
=\int_{0}^{R}{\rm d}r\,r^{3}{\cal L}_{\rm{E}}(r)$\\
\resizebox{7.1cm}{!}{\includegraphics{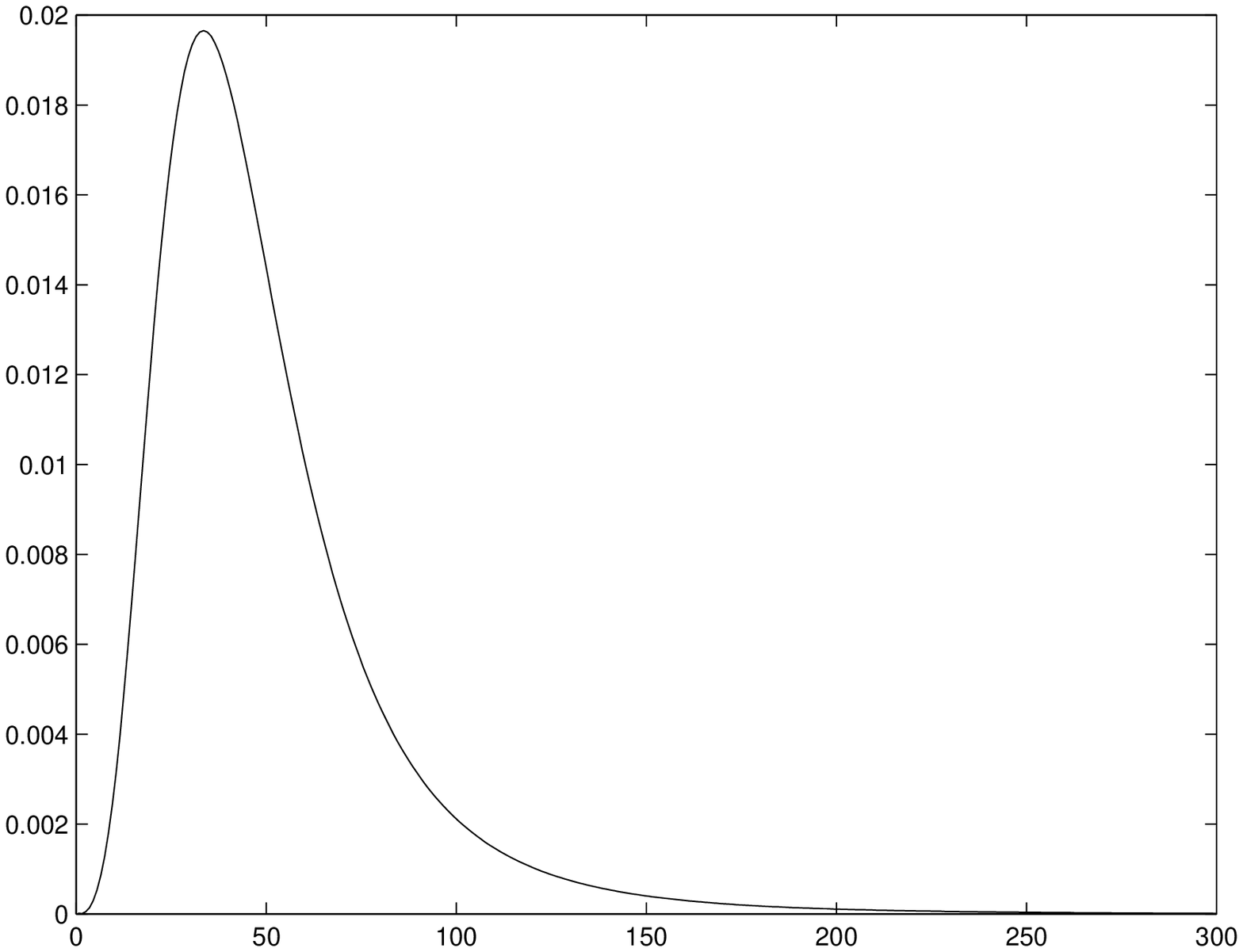}}r
~~~~&~~~~\resizebox{7.1cm}{!}{\includegraphics{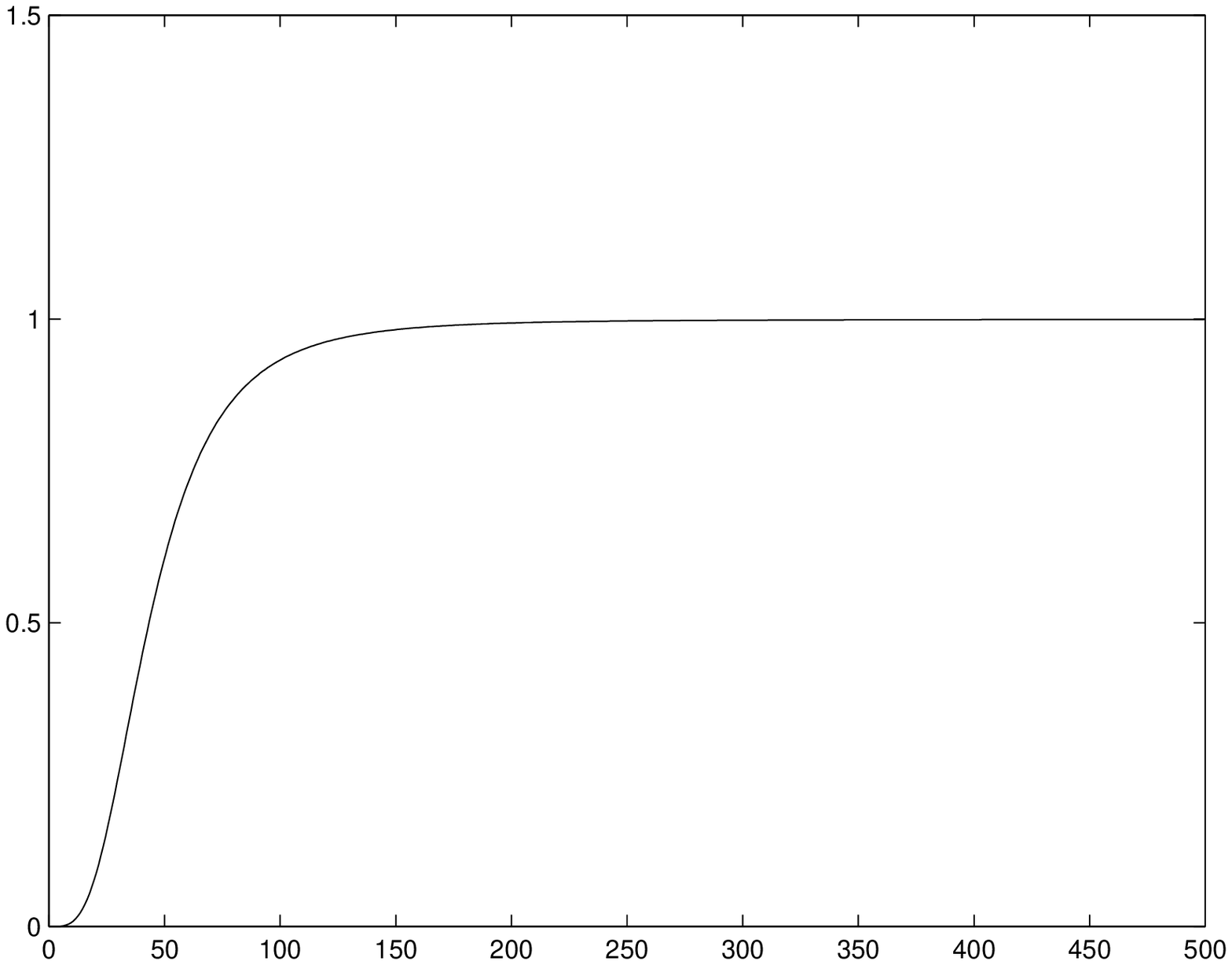}}R\\
~~~~~~~~~~~~Fig.~2. \textsc{Lagrangian}~~~~
&~~~~~~~~~~~~~~~~~~\,\,\,Fig.~3. \textsc{Action}~~~~
\end{tabular}

\section{Comment}
In four-dimensions the Bianchi identity (\ref{Bianchi}) 
and  the equation of motion~(\ref{eqm}) show that 
 ${{}^{\ast} F}^{\mu\nu}$ and $\Pi^{\mu\nu}$ satisfy 
the same equation. From this observation one  may 
consider solutions of 
\begin{equation}
\Pi^{\mu\nu}=c\,{{}^{\ast} F}^{\mu\nu}
\label{self}
\end{equation}
where $c$ is a constant.\newline
Eq.(\ref{self}) reduces to the ordinary self-dual equation for the 
Yang-Mills action. Hence, one may regard eq.(\ref{self}) as a 
self-dual equation for general actions.\newline
In the case of $\mbox{SU}(2)$, from  the completeness
relation~(\ref{complete}) and eq.(\ref{Pi}),   
eq.(\ref{self}) leads to
\begin{equation}
0=(G+\kappa F)^{-1}{}^{\mu}{}_{\alpha}{}^{\nu}{}_{\beta}
-\textstyle{\frac{1}{2}}(G+\kappa
F)^{-1}{}^{\mu}{}_{\gamma}{}^{\nu}{}_{\gamma}\delta_{\alpha\beta}
+\kappa c\,(\Det(G+\kappa F))^{-\frac{1}{4}}\,
{{}^{\ast} F}^{\mu}{}_{\alpha}{}^{\nu}{}_{\beta}
\end{equation}
Using the identity 
\begin{equation}
F^{a}_{\mu\nu}{{}^{\ast} F}^{a\nu\lambda}=-\textstyle{\frac{1}{4}}
F^{a}_{\nu\rho}{{}^{\ast} F}^{a\nu\rho}\,\delta_{\mu}{}^{\lambda}
\end{equation}
we find that 
in the case of $\mbox{SU}(2)$  eq.(\ref{self})  is  equivalent to
\begin{equation}
\left((\Det(G+\kappa
F))^{\frac{1}{4}}-\textstyle{\frac{1}{8}}c\kappa^{2}
\tr(F_{\lambda\rho}{{}^{\ast} F}^{\lambda\rho})\right)F_{\mu\nu}=
c\,{{}^{\ast} F}_{\mu\nu}+\textstyle{\frac{1}{2}}c\kappa
[F_{\mu}{}^{\lambda},{{}^{\ast} F}^{\lambda}{}_{\nu}]
\label{selfdual}
\end{equation}
However, 
direct calculation  shows   that the $\mbox{SO}(4)$ invariant ansatz 
(\ref{ansatz}) is not compatible with eq.(\ref{selfdual}). 
\newline
\newline
\begin{center}
\large{\textbf{Acknowledgements}}
\end{center}
The author  wishes to 
thank T. Ahn,  H. Osborn, M. Green and A. Mountain  at DAMTP,  
University of Cambridge for  valuable discussions.  
This work was partly supported by Cambridge Overseas Trust and  
Asia Pacific Center for Theoretical Physics. 
\newline
\newline

\newpage
\appendix
\begin{center}
\Large{\textbf{Appendix}}
\end{center}
\renewcommand{\theequation}{A.\arabic{equation}} 
The linearization of the  differential
equation~(\ref{eqmsu2}) around the stationary solutions, \newline  
$f=0,1,\frac{1}{2}$ may be achieved by writing $f(r)\simeq 
f+\delta f(r)$. \newline
Around $f=0,1$ we have
\begin{equation}
\displaystyle{
0=\delta f^{\prime\prime}+\frac{1}{r}\delta f^{\prime}-\frac{4}{r^{2}}\delta f}
\end{equation}
The  solutions  are  $\delta f=r^{\pm 2}$, 
which  are   consistent with  
eqs.(\ref{instanton},\,\ref{instanton2}).\newline
Around $f=\frac{1}{2}$ we have
\begin{equation}
\displaystyle{
0=(1-\kappa\frac{1}{2r^{2}})(1+\kappa\frac{1}{r^{2}})\delta f^{\prime\prime}+
(\frac{1}{r}+\kappa\frac{5}{2r^{3}})\delta f^{\prime}+(\frac{2}{r^{2}}
+\kappa\frac{2}{r^{4}})\delta f}
\end{equation}
The asymptotic solution  is  
\begin{equation}
\begin{array}{cc}
f(r)\simeq \textstyle{\frac{1}{2}}+c\sin (\sqrt{2}\ln
(r/r_{\rm{o}}))~~~~&~~~~
\mbox{for}~r>>1
\end{array}
\label{half}
\end{equation}
where $c,r_{\rm{o}}$ are free parameters.

\newpage
\bibliographystyle{unsrt}
\bibliography{reference}

\begin{thebibliography}{10}

\bibitem{Born-Infeld}
{M}. {B}orn and {L}. {I}nfeld.
\newblock {Foundations of the New Field Theory}.
\newblock {\em Proc. Roy. Soc.}, A144:~425, 1934.

\bibitem{JPA143059}
{T}. {H}agiwara.
\newblock {A non-abelian Born-Infeld Lagrangian}.
\newblock {\em J. Phys.}, A14:3059, 1981.

\bibitem{NPB330151}
{{P}. {C}. {A}rgyres and {C}. {R}. {N}appi}.
\newblock {Spin-1 Effective Actions from Open Strings}.
\newblock {\em Nucl. Phys.}, B330:~151, 1990.

\bibitem{NPB50141}
{{A}. {A}. {T}seytlin }.
\newblock {On non-Abelian Generalisation of Born-Infeld Action in String
  Theory}.
\newblock {\em Nucl. Phys.}, B501:~41, 1997.

\bibitem{NPB81118}
{{J}. {S}cherk and {J}. {H}. {S}chwarz}.
\newblock {Dual Models for Non-Hadrons}.
\newblock {\em Nucl. Phys.}, B81:~118, 1974.

\bibitem{NPB276391}
{A}.~{A}. {T}seytlin.
\newblock {Vector Field Effective Action in the Open Superstring Theory}.
\newblock {\em Nucl. Phys.}, B276:~391, 1986.

\bibitem{9801127}
{{D}. {B}recher and {M}. {J}. {P}erry}.
\newblock {Bound States of D-Branes and the Non-Abelian Born-Infeld Action}.
\newblock hep-th/9801127.

\bibitem{9804180}
{{D}. {B}recher}.
\newblock {BPS States of the Non-Abelian Born-Infeld Action}.
\newblock hep-th/9804180.

\bibitem{9901073}
{N. Grandi, E. F. Moreno and F. A. Schaposnik}.
\newblock {Monopoles in non-Abelian Dirac-Born-Infeld Theory}.
\newblock hep-th/9901073.

\bibitem{PLB5985}
{{A}. {A}. {B}elavin, {A}. {M}. {P}olyakov, {A}. {S}. {S}chwartz and {Y}u. {S}.
  {T}yupkin}.
\newblock {Pseudoparticle Solutions of the Yang-Mills Equations}.
\newblock {\em Phys. lett.}, 59B:~85, 1975.

\bibitem{weinberg23}
{{S}. {W}einberg}.
\newblock {\em {The Quantum Theory of Fields}}.
\newblock Cambridge University Press, 1996.
\newblock See chapter 23 and references therein.

\end{thebibliography}
\end{document}